\documentstyle[12pt,aps,epsfig,prl]{revtex}

\relax  
\begin{document}
\preprint{prl}
\draft

\title{Josephson plasma resonance in $\kappa-$(BEDT-TTF)$_2$Cu(NCS)$_2$}

\bigskip

\author{M. M. Mola, J. T. King, C. P. McRaven and S. Hill\cite{email}}
\address{Department of Physics, Montana State University, Bozeman, MT 59717}

\author{J. S. Qualls and J. S. Brooks}
\address{Department of Physics and National High Magnetic Field Laboratory, Florida State University,
Tallahassee, FL 32310}

\date{\today}
\maketitle

\bigskip

\begin{abstract}
A cavity perturbation technique is used to study the microwave
response of the organic superconductor
$\kappa-$(BEDT-TTF)$_2$Cu(NCS)$_2$. Observation of a Josephson
plasma resonance, below T$_c$ ($\sim$10 K), enables investigation
of the vortex structure within the mixed state of this highly
anisotropic, type$-$II, superconductor. Contrary to previous
assumptions, frequency dependent studies (28 $-$ 153 GHz) indicate
that the squared plasma frequency ($\omega_p^2$) depends
exponentially on the magnetic field strength. Such behavior {\em
has} been predicted for a weakly pinned quasi-two-dimensional
vortex lattice $[$Bulaevskii {\em et al.} Phys. Rev. Lett. {\bf
74}, 801 (1995)$]$, but has not so far been observed
experimentally. Our data also suggests a transition in the vortex
structure near the irreversibility line not previously reported
for an organic superconductor using this technique.

\end{abstract}

\smallskip


\bigskip

It is widely acknowledged in the layered high$-$T$_c$
superconductors (HTSs) and, more recently, in the layered organic
superconductors (OSs), that Josephson coupling is responsible for
interlayer transport of superconducting pairs
\cite{one,oneb,two,three,four}.  Magneto-optical experiments have
been carried out by many groups, and for a wide range of
materials, {\em e.g.} Bi$_2$Sr$_2$CaCu$_2$O$_{8+\delta}$ (BSCCO)
\cite{one,two} YBa$_2$Cu$_3$O$_{6-y}$ (YBCO) \cite{two}, as well
as several OSs \cite{three,four}. In each case, these experiments
have demonstrated that a Josephson plasma resonance (JPR) is
observed in the inter-layer conductivity, and that this resonance
may be used as an extremely sensitive tool for probing the vortex
structure/dynamics.

$\kappa-$(BEDT-TTF)$_2$Cu(NCS)$_2$ (BEDT-TTF denotes
bis-ethylenedithio-tetrathiafulvalene, or ET for short) is one in
a series of highly anisotropic OSs, characterized by a layered
structure. The conducting layers consist of a checkerboard pattern
of face-to-face ET dimers, with nearest neighbor dimer pairs
oriented perpendicularly \cite{fourb}. These donor sheets exhibit
near isotropic 2D conductivity within the {\bf bc} plane. The
insulating layers form from weakly bonded arrays of $V-$shaped
Cu(NCS)$_2$ anions. The anisotropy parameter in the normal state,
given by the ratio of the in-plane to out-of-plane conductivities
$\sigma_{bc}/\sigma_a$, is $\sim 1000$. In the superconducting
state, the anisotropy parameter is given by $\gamma$ $\equiv$
$\lambda_\perp/\lambda_\parallel \sim 100-200$, where $\lambda
_\parallel$ and $\lambda _\perp$ are the London penetration depths
for AC currents induced parallel ($\lambda _\parallel \sim 0.8 \mu
m$) and perpendicular ($\lambda _\perp \sim 100 \mu m$) to the
conducting layers respectively \cite{five}. Such a large
anisotropy makes this OS a prime candidate to study Josephson
coupling, and changes in this coupling upon the introduction of
supercurrent vortices into the sample through the application of
an external magnetic field.

In highly anisotropic superconductors, it is expected that the
plasma mode for the low conductivity direction ($\sigma_a$) lies
below the superconducting gap {\em i.e.} $\hbar \omega_p <
2\Delta$. Below the critical temperature T$_c$, this plasma mode
dominates the {\bf a}-axis microwave response with frequency,
$\omega_p$, which depends on the maximum inter-layer (or
Josephson) current J$_m$({\em B},T), through the expression


\[
\omega _p^2 (B\text{,T}) = \frac{{8\pi ^2 cs}} {{\varepsilon _c
\Phi _0 }}\text{J}_m (B\text{,T}),
\]

\begin{tabbing}
\noindent{where} \`{(1)}
\end{tabbing}

\centerline{J$_m$ = J$_o \langle\langle \cos\varphi_{n,n+1}$({\bf
r})$\rangle _t \rangle _d$,}
\bigskip

\noindent {$s$ is the crystal inter-layer spacing, $\varepsilon_c$
is the high frequency permittivity, $\Phi_o$ is the flux quantum,
$\varphi_{n,n+1}({\bf r})$ is the gauge-invariant phase difference
between layers $n$ and $n$ + 1 at a point {\bf r} = {\em x,y} in
the {\bf bc} (high conductivity) plane, and $\langle....\rangle_t$
and $\langle....\rangle_d$ denote thermal and disorder averages.
J$_o$(T) = $c\Phi_o/8\pi^2s \lambda_{\perp}^2$(T) is the maximum
inter-layer Josephson current density at zero field (B$_{DC}$ =
0), and $\lambda_{\perp}$ is the inter-layer London penetration
depth \cite{six,seven}.}

An important property of the Josephson coupling is the influence
of an externally applied magnetic field on the collective plasma
oscillation frequency $\omega_p$. When the applied DC magnetic
field is parallel to the least conducting axis, a mixed state is
created in which the field penetrates the sample in quantized flux
tubes, generating supercurrent vortices in the superconducting
layers. $\varphi_{n,n+1}({\bf r})$ depends explicitly on the
vortex structure within this mixed state and is, thus, responsible
for the field dependence of the resonance frequency $\omega_p$
\cite{six}. If the flux tubes form straight lines along the {\bf
a}-axis, $\langle \cos \varphi_{n,n+1}({\bf r}) \rangle$ = 1, and
maximum Josephson coupling occurs. However, in the presence of
disorder, {\em e.g.} as a result of crystal defects which create
vortex pinning sites, or through thermal fluctuations, the flux
tubes will deviate from straight lines. This suppresses the
maximum Josephson current. It is when AC currents are excited
between the layers, at a frequency which corresponds to the
natural frequency of the plasma oscillation ($\omega_p$), that a
sharp resonance is observed. Consequently, this resonance
frequency provides a direct measurement of the maximum inter-layer
current density, which in turn can be used to probe vortex
structure in the mixed state.

We have carried out experiments on several
$\kappa-$(ET)$_2$Cu(NCS)$_2$ single crystals. The dimensions of
each crystal were approximately $0.75 \times 0.5 \times 0.2$
mm$^3$, with the low conductivity axis the shortest of the three.
We found that all samples gave qualitatively similar results.
Microwave impedance measurements were carried out using a cavity
perturbation technique described elsewhere \cite{eight}. We used
the TE01$n$ modes of cylindrical copper cavities ($n$ = 1, 2, 3,
{\em etc.}), and the TE112 mode of a rectangular copper cavity.
The use of a range of cavities enabled wide frequency coverage
from $28-153$ GHz. Typical loaded Quality factors for these
cavities for a given measurement were approximately $5 \times
10^3$ to $2.5 \times 10^4$. The samples were placed in one of two
configurations within the cavity, which correspond to two
different microwave field configurations \cite{eight}. The first
position induces strictly in-plane currents, with the AC microwave
magnetic field perpendicular to the layers (H$_{AC}$ $\perp$ {\bf
bc}). The second configuration, with H$_{AC}$ $\parallel$ {\bf
bc}, generates both in-plane and inter-layer Josephson currents;
for a discussion of the electrodynamics of organic
superconductors, see ref \cite{lambda}. Both cavity positions were
such that the DC magnetic field (B$_{DC}$) was perpendicular to
the conducting planes, {\em i.e.} B$_{DC}$ $\parallel$ {\bf a}.
Temperature control was achieved using a small resistive heater
attached mechanically to the cavity, and a Cernox thermometer. For
inter-layer measurements, DC magnetic fields were generated using
an 8 T superconducting solenoid. The in-plane measurements were
conducted in the 33 T resistive magnets at the National High
Magnetic Field Laboratory in Tallahassee, FL. In both cases, field
sweeps were made at a rate of approximately 1 T per minute.

Figure 1 shows dissipation due to in-plane currents, plotted
versus magnetic field, for temperatures in the range 1.6 to 4.5 K;
the frequency is 44.4 GHz and the traces are offset for clarity.
As expected for this configuration, the dissipation ($\propto$
surface resistance) increases monotonically with increasing DC
field. Indeed, the surface resistance approximately follows a
B$_{DC}^{1/2}$ behavior, consistent with flux-flow type
dissipation \cite{nine}. It is interesting to note the lack of (or
extremely weak) temperature dependence in the various traces in
Fig. 1 \cite{SdH}. This is unexpected given that $\mu_o$H$_{c2}$
($\sim$ $2-3$ T at 2 K) varies by at least 0.5 T over the
equivalent temperature range. At present, the origin of this
behavior is not known and will form the basis for future
investigations. The main point to note from the in-plane
measurements is the lack of any resonant features (see Fig. 2
below). These findings are in stark contrast to recent studies by
Schrama {\em et al.} \cite{schrama}, where it is claimed that an
observed resonance is attributable to the in-plane response.


The second configuration, in which a mixture of in-plane and
inter-layer currents are exited ({\em i.e.} H$_{AC}
\parallel$ {\bf bc}), gives very different results from those shown in Fig. 1.
Fig. 2 plots dissipation versus magnetic field for several
frequencies in the range $28-153$ GHz; the temperature is 2.0 K in
each case. A pronounced resonant feature can now be seen
(indicated by arrows), which becomes sharper and moves to lower
magnetic field upon increasing the frequency. Due to the absence
of this resonance in the purely in-plane response (Fig. 1), we
conclude that this resonance is related to dissipation within the
sample caused by inter-layer currents $[i.e.$ dissipation $\propto
\sigma_a (\omega)]$, and that insufficient care was taken to
separate these contributions in the work of Schrama {\em et al.}
\cite{schrama,comment}. The anti-cyclotronic nature of the
resonances shown in Fig. 2 has been well documented in the HTSs
\cite{ten}, and has also been observed in previous work on this OS
\cite{three}, though only two frequencies were used. The wider
frequency range employed here clearly confirms the
anti-cyclotronic trend.

Figures 3a and b show the temperature dependence of the resonance
for $\omega_p/2\pi$ = 76 and 111 GHz, respectively. Note that, as
the temperature increases toward the superconducting to normal
transition (T$_c$ $\sim$ 10 K), the resonances vanish. This
observation would seem to indicate that it is the superconducting
carriers that are responsible for the sharp dissipative peak. This
fact, along with the anti-cyclotronic nature of the resonance, and
the AC current polarization dependence of the effect ({\em i.e.}
resonances are not observed for H$_{AC} \perp$ {\bf bc}), leads us
to conclude that the resonance indeed corresponds to a Josephson
plasma mode. Note, no apparent hysteresis was observed for the up
and down sweeps of the magnetic field in this material, hence only
up sweeps are plotted for clarity.

Looking more closely at the 76 GHz data in Fig. 3a, we see that
the resonance peak position decreases monotonically with respect
to increasing temperature. This behavior is consistent with
previous observations in this OS \cite{three}. However, we see an
entirely different temperature dependence in the tail of the
resonance $-$ the structure in the low field tail ($\sim$ 0.1 T)
shows an initial increase in field with increasing temperature,
reaching a maximum, then decreasing back to zero field as T$_c$ is
approached. This so-called "cusp" behavior has been observed in
some HTSs \cite{ten}, but never in any of the OSs.

For the 111 GHz data in Fig. 3b, we see a similar resonance
structure, but now the cusp behavior can be seen in the position
of the resonance rather than its tail. The phenomenon responsible
for the cusp has not changed, however, the JPR has been pushed to
lower field by working at higher frequencies, thereby moving the
resonance position into the cusp regime.

Measurements spanning a broad frequency range, as shown in Fig. 4,
indicate that the cusp behavior is confined to a relatively narrow
field and temperature domain. In fact, comparison of the cusp
position with published magnetization data \cite{eleven,twelve}
show that it falls along, or very near to, the irreversibility
line. Due to the change in nature of the temperature dependence of
the JPR at the cusp position, we conclude that there is a
transformation in the vortex structure at (or near) the
irreversibility line.

It is well known that a dimensional crossover phenomenon between a
three-dimensional flux-line-lattice and a quasi-two-dimensional
(Q2D) pancake vortex lattice occurs at approximately 10 mT in
$\kappa-$(ET)$_2$Cu(NCS)$_2$ \cite{thirteena,thirteen}. Thus, it
is assumed that, over a large portion of the superconducting phase
diagram probed in this work (and shown in Fig. 4), there is strong
disorder in the pancake positions as viewed along the {\bf a}-axis
(least conducting direction). At the same time, however, long
range order may be present within the superconducting layers ({\em
i.e.} a Q2D vortex lattice) below some characteristic melting
temperature [T$_m$({\em B})]. We propose two possible explanations
for the cusp phenomenon, both of which involve transitions in the
vortex structure, and both of which account qualitatively for the
observed {\em B},T-dependence of the plasma frequency.


One possible explanation for the cusp phenomenon involves a 2D
solid to liquid melting transition \cite{fourteen,fifteen}. This
occurs when thermal (or quantum) fluctuations in the positions of
the vortices become comparable to the inter-vortex separation,
{\em i.e.} either when the flux density exceeds a critical value
{\bf B}$_m$(T), or when the temperature exceeds a critical value
T$_m$({\bf B}). Below {\bf B}$_m$(T) and T$_m$({\bf B}), in the
Q2D pancake vortex lattice phase, random defects will tend to pin
each Q2D solid in such a way that there is little correlation
between the locations of vortices in adjacent layers, {\em i.e.}
$\langle \cos \varphi_{n,n+1} \rangle$ will be small, and
inter-layer Josephson tunneling will be suppressed. Upon
increasing the temperature slightly, thermal fluctuations will
tend to cause the vortices in each layer to deviate from their
mean positions. Provided that these displacements are small
compared to the distance between adjacent vortices, the effect of
increasing temperature will actually result in an increase in the
inter-layer tunneling probability and, hence, an increase in the
maximum inter-layer Josephson current. Eventually, upon further
increasing the temperature, the vortex lattice melts. In the
liquid phase, increased thermal fluctuations ({\em i.e.}
increasing temperature) will tend to suppress inter-layer
tunneling and, hence, the maximum inter-layer Josephson current.
Thus, completely opposite temperature dependences of the JPR
frequencies are expected in the Q2D solid and liquid phases (see
Eq. 1), {\em i.e.} exactly as observed experimentally.

Another possible explanation for the cusp phenomenon involves a
depinning transition \cite{depin1,depin2} whereby, upon increasing
the temperature, thermal fluctuations exceed some critical
depinning threshold. As this threshold is approached, one can
think of a "pinned" Q2D vortex lattice performing larger and
larger collective displacements from equilibrium, until eventually
it becomes completely "depinned," or mobile. In this sense,
lattice melting and depinning represent qualitatively similar
phenomena $-$ indeed, under certain conditions, lattice melting
and depinning may occur together \cite{depin2}. However, the
physical origin of each transition is very different: in the case
of melting, thermal fluctuations overcome the inter-vortex
interactions which maintain long range Q2D order; in the case of
depinning, thermal fluctuations suppress the influence of random
defects, and long range Q2D ordering of the vortices persists. For
both cases, one should expect opposing temperature dependences for
$\langle \cos \varphi_{n,n+1} \rangle$ above and below the
respective transitions, since it is pinning that is responsible
for the suppression of interlayer tunneling at low temperatures,
while the effects of this pinning are absent, or at least greatly
diminished (see below), in the high temperature liquid or depinned
states.

We can now relate the observed magnetic field and temperature
dependence of the resonances to the above discussion. Application
of a magnetic field tends to reduce the JPR frequency
$\omega_p$({\bf B},T) (see discussion below). Thus, for a fixed
frequency measurement, a resonance is observed when
$\omega_p$({\bf B},T) matches the measurement frequency.
Consequently, any trend that increases $\omega_p$({\bf B},T) will
shift the observed resonance to higher magnetic fields, since a
stronger field is required to reduce $\omega_p$({\bf B},T) so that
it matches the measurement frequency. The converse is true for
trends which decrease $\omega_p$({\bf B},T). Thus, in the pinned
Q2D solid phase, increasing temperature will tend to shift the
resonance to higher field while, in the liquid or depinned Q2D
solid phase, one expects the resonance to move to lower fields
upon increasing the temperature. This is precisely the behavior
observed in Fig. 4b, where the cusp marks the change in
temperature dependence of the resonance.

Next, we consider the frequency dependence of the resonance (at
constant temperature). It has previously been assumed for this
substance that the frequency of the resonance ($\omega_p$) obeys a
power-law field dependence of the form

\bigskip

\begin{tabbing}
\centerline{$\omega_p^2$ = A{\em B}$^{-\mu} \exp$(T/T$_c$),}
\`\hspace{-5cm} (2)
\end{tabbing}

\bigskip

\noindent{where A is a constant. The inset to Fig. 5 shows two
fits to a power-law for T = 2.0 K: the dashed line is a fit to
data covering a wide field range; the dotted line is a fit to data
above the melting/depinning transition only. The former fit gives
a value for the exponent of $\mu=0.7$, which is consistent with
previous work where only two data points were available
\cite{three}. The latter fit gives a value for the exponent of
$\mu=1.7$. Overall, the power-law does not appear to be a good
fit, even when restricted to data above the melting/depinning
transition, {\em i.e.} closer inspection of the inset to Fig. 5
shows that the exponent $\mu$ increases with increasing field. We
also note that an exponent $\mu>1$ is unphysical based on the
various theories which predict a power-law field dependence of the
squared plasma frequency \cite{seven,kbm}; $\mu$ of order unity
corresponds to little or no correlation between vortices in
adjacent layers, while $\mu<1$ indicates some degree of ordering
of the vortices along lines parallel to the {\bf a}-axis. In the
limiting case of a vortex liquid, $\mu=1$ has been predicted
\cite{kbm}, and $\mu$ close to unity {\em has} been observed
experimentally in BSCCO \cite{one,oneb}.

In the main panel of Fig. 5, we show two further fits to the data
using a week pinning theory developed by Bulaevskii {\em et al.}
\cite{six}, where a decaying exponential field dependence of
$\omega_p^2$({\bf B},T) is predicted; the dashed and dotted curves
correspond to the same field ranges as the power-law fits
discussed above. In this case,}

\bigskip


\[
\omega _p^2 (B\text{,T}) = \frac{{8\pi ^2 cs}} {{\varepsilon _c
\Phi _0 }}\text{J}_0 \exp [ - (B/B_D
)^{{\raise0.5ex\hbox{$\scriptstyle 3$} \kern-0.1em/\kern-0.15em
\lower0.25ex\hbox{$\scriptstyle 2$}}} ],
\]

\begin{tabbing}
\noindent{where} \`{(3)}
\end{tabbing}


\[
B_D  = \frac{{\Phi _0^{{\raise0.5ex\hbox{$\scriptstyle {11}$}
\kern-0.1em/\kern-0.15em \lower0.25ex\hbox{$\scriptstyle 3$}}} }}
{{(4\pi )^3 (2\pi \gamma _0 E_p )^{{\raise0.5ex\hbox{$\scriptstyle
2$} \kern-0.1em/\kern-0.15em \lower0.25ex\hbox{$\scriptstyle 3$}}}
s^2 \lambda _\parallel ^4 }}.
\]

\bigskip

\noindent{B$_D$ is a de-coupling field, above which inter-layer
phase coherent tunneling ceases \cite{Bd}; $\gamma_o$ is the
anisotropy parameter; E$_p$ is a pinning parameter; and
$\lambda_{\parallel}$ is the in-plane penetration depth
\cite{seven}. Clearly, the above expression yields superior fits
to the data over both field ranges (there is only a $\sim 10\%$
difference in the fitting parameters obtained from the dashed and
dotted curves).

At the present time it is not clear why the Bulaevskii theory,
which was developed for a weakly pinned vortex solid, fits our
data so well. In particular, since most of the data in Fig. 5 were
obtained above the melting/depinning transition, one might expect
a 1/{\em B} dependence of $\omega_p^2$, as predicted for a vortex
liquid phase \cite{kbm}. One possible conclusion is that
quasi-long range 2D order persists in this high temperature
regime, and that pinning continues to play a role. Further
detailed frequency dependent studies will be required to resolve
this issue.

It is interesting to note that, if one adopts the weak pinning
theory, it is possible to extract certain sample parameters for
comparison with published results. Using $\varepsilon_c$ = 25,
$\gamma_o$ = 100$-$200, and $\lambda_{\parallel}$ =1 $\mu$m
\cite{five}, we find the zero field maximum interlayer current
density, J$_o$ = $2 \times 10^3$ A cm$^{-2}$. From this, we are
able to deduce the interlayer penetration depth, $\lambda_{\perp}$
= 90 $\mu$m, which is on the order of previous measurements
\cite{five}. Further, we find a value for the decoupling field,
B$_D$ = 0.7 T, from which we can extract the pinning parameter
E$_p$ = $6.3 \times 10^{-13}$ to $1.3 \times 10^{-12}$
J$^2$m$^{-4}$. This range of values is three orders of magnitude
lower than the BSCCO measurements discussed in ref.
\cite{sixteen}. This is not surprising, since it is well
established that the ET salts are considerably cleaner systems
than the HTSs, thus having far fewer pinning sites. Moreover, for
the weak pinning limit, it has been noted that the in-plane
critical current depends on the pinning parameter given above
\cite{sixteen}:}

\bigskip

\begin{tabbing}
\centerline{J$_{ab}$ = ($4c$/3$\Phi_o$)(E$_p$/3)$^{1/2}$.}
\`\hspace{-5cm} (4)
\end{tabbing}

\bigskip

\noindent{Using the range of values above for E$_p$, we find the
critical in-plane current density to be in the range, J$_{ab}$ =
$3 \times 10^4$ A cm$^{-2}$ to  $4 \times 10^4$ A cm$^{-2}$,
consistent with values obtain by other methods \cite{three}. Thus
our measurements have proven consistent with previous work, and
seem to agree excellently with the weak pinning theory of
Bulaevskii {\em et al.} \cite{six}.

Finally, we have noted a slight sample dependence in our
measurements, as well as some variance compared to other published
results \cite{three,schrama}. The most notable differences can be
found in the resonance field positions for a given frequency and
temperature, as is clearly seen in Fig. 4, where several of the
curves corresponding to different frequencies lie on top of each
other and even cross in one case. This sample dependence is,
perhaps, not surprising given the dependence of $\omega_p$({\bf
B},T) on the pinning parameter, E$_p$, which is liable to depend
on sample quality. However, we also see an apparent scatter in the
cusp positions in Fig. 3. At this stage, it is unclear whether
this scatter is real, {\em i.e.} attributable to sample quality,
or whether it is associated with uncertainties introduced during
data analysis. Establishing whether there is a sample dependence
in the cusp positions could help resolve the nature of the
proposed transition in the vortex structure/dynamics; a first
order melting transition should not be influenced by sample
quality, whereas one would expect a depinning transition to
exhibit pronounced sample dependence (see refs. \cite{four,inada},
for recent results on the $\kappa$-phase ET salts).

In conclusion, Josephson plasma resonance measurements have
provided a wealth of information regarding the superconducting
properties of the organic charge transfer salt
$\kappa-$(ET)$_2$Cu(NCS)$_2$. Due to the extremely clean nature of
this material, it is apparent that the power-law field dependence
of the squared plasma frequency, as observed in several
High$-$T$_c$ compounds, is inapplicable in the present case.
Instead, we are able to fit our data with a weak pinning theory,
from which we obtain excellent agreement with several well known
material parameters, {\em e.g.} the inter-layer penetration depth,
and the in-plane critical current. This novel technique also
allows us to detect subtle changes in the Q2D vortex
structure/dynamics. In particular, we see a cusp in the
temperature dependence of the plasma frequency $\omega_p$({\bf
B},T) which may be due to either vortex lattice melting or
depinning. Further examination of this cusp should lead to a
better understanding of the observed transition, and to the
properties of vortex matter in general.

This work was supported by the Petroleum Research Fund (33727-G3)
and the Office of Naval Research (N00014-98-1-0538). CPM and JTK
acknowledge support through the NSF REU program (NSF-DMR 9820388).


\clearpage

\noindent{{\bf Figure captions}}

\bigskip

\noindent{Fig. 1. Field dependent dissipation ($\propto$ surface
resistance) due to in-plane currents, at various temperatures, and
for $\omega$/2$\pi$ = 44.4 GHz (offset for clarity). As expected
for flux flow type dissipation, the surface resistance varies
approximately as B$^{1/2}$ for B$<\mu_o$H$_{c2}$. This
configuration shows no resonant behavior and little temperature
dependence. The inset depicts the geometry of the crystal, the
orientations of the applied DC and AC magnetic fields, and the AC
current paths around the edges of the sample.}

\bigskip

\noindent{Fig. 2. Magnetic field sweeps, at various frequencies,
with H$_{AC}$ $\parallel$ {\bf bc}; the temperature is 2.0 K in
each case. Notice the sharp resonance feature, indicated by
arrows, which is attributable to inter-layer Josephson currents.
Note that the resonance field position, B$_o$, decreases with an
increasing frequency, {\em i.e.} anti-cyclotronic behavior.}

\bigskip

\noindent{Fig. 3. Temperature dependence of the Josephson plasma
mode for a) $\omega$/2$\pi$ = 76 GHz and b) $\omega$/2$\pi$ = 111
GHz. Notice that in a), the resonance peak position decreases
monotonically as a function of temperature; however, the structure
in the tail of this resonance exhibits a "cusp" behavior. In
contrast, due to the anti-cyclotronic behavior of the JPR, the
resonance in b) has moved to lower field, and we now see a cusp
behavior in the resonance position. The inset depicts the geometry
of the crystal, the orientations of the applied DC and AC magnetic
fields, and the AC current paths across the surfaces of the
sample.}

\bigskip

\noindent{Fig. 4. a) Plot of the cusp positions in field, versus
temperature, for several samples and frequencies; the inset plots
several representative series of resonance fields (for a single
sample and a range of frequencies) versus temperature, and the
method used to determine the location of the cusp, {\em i.e.} the
crossing points of the extrapolated fits (dashed lines) to the low
and high temperature behavior. The cusp positions in the main
panel of a) are scattered around the irreversibility line (solid
line). b) Resonance field B$_o$ vs. temperature for measurements
spanning a wide frequency range (28-153 GHz), and for several
samples. The higher frequencies exhibit a cusp behavior, while
lower frequencies show a monotonic temperature dependence. The
dashed line indicates the approximate position of irreversibility
line, and the solid lines are guides to the eye.}

\bigskip

\noindent{Fig. 5. When $\omega_p^2$ is plotted against B$^{3/2}$,
an exponential decay is observed. This behavior is predicted for a
weakly pinned vortex lattice \cite{six}. The dashed line is a fit
to all of the data points, while the dotted line represents a fit
to data above the melting/depinning transition. Inset: we have
tried to fit the data with a power-law, as has been observed in
the HTSs, and previously claimed in this material [3]. However,
this is clearly not a good fit when a wide field/frequency
interval is sampled. The dashed line represents a fit to all of
the data, giving an exponent $\mu=0.7$, while the dotted line is a
fit to data above the melting/depinning transition, giving
$\mu=1.7$.}

\clearpage

\begin{figure}
\centerline{\epsfig{figure=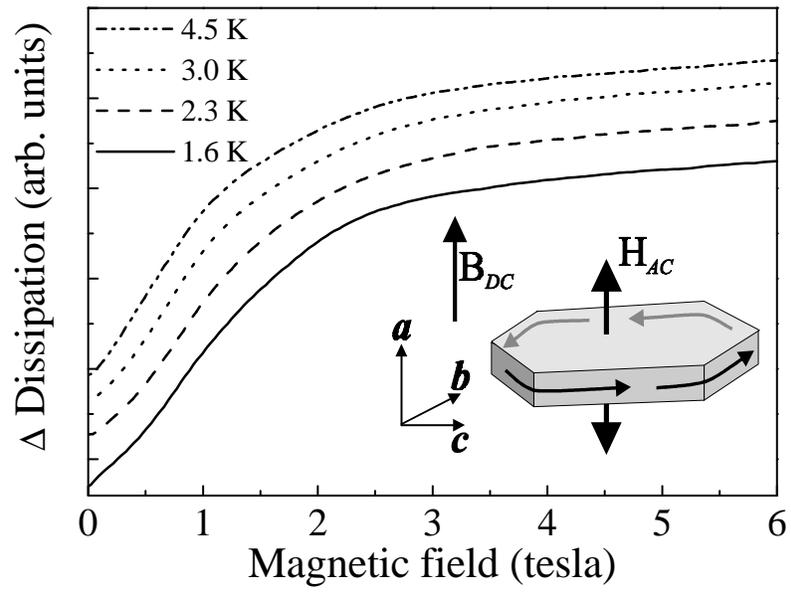,width=120mm}}
\bigskip \caption{M. Mola
{\em et al.}}

\end{figure}

\clearpage

\begin{figure}
\centerline{\epsfig{figure=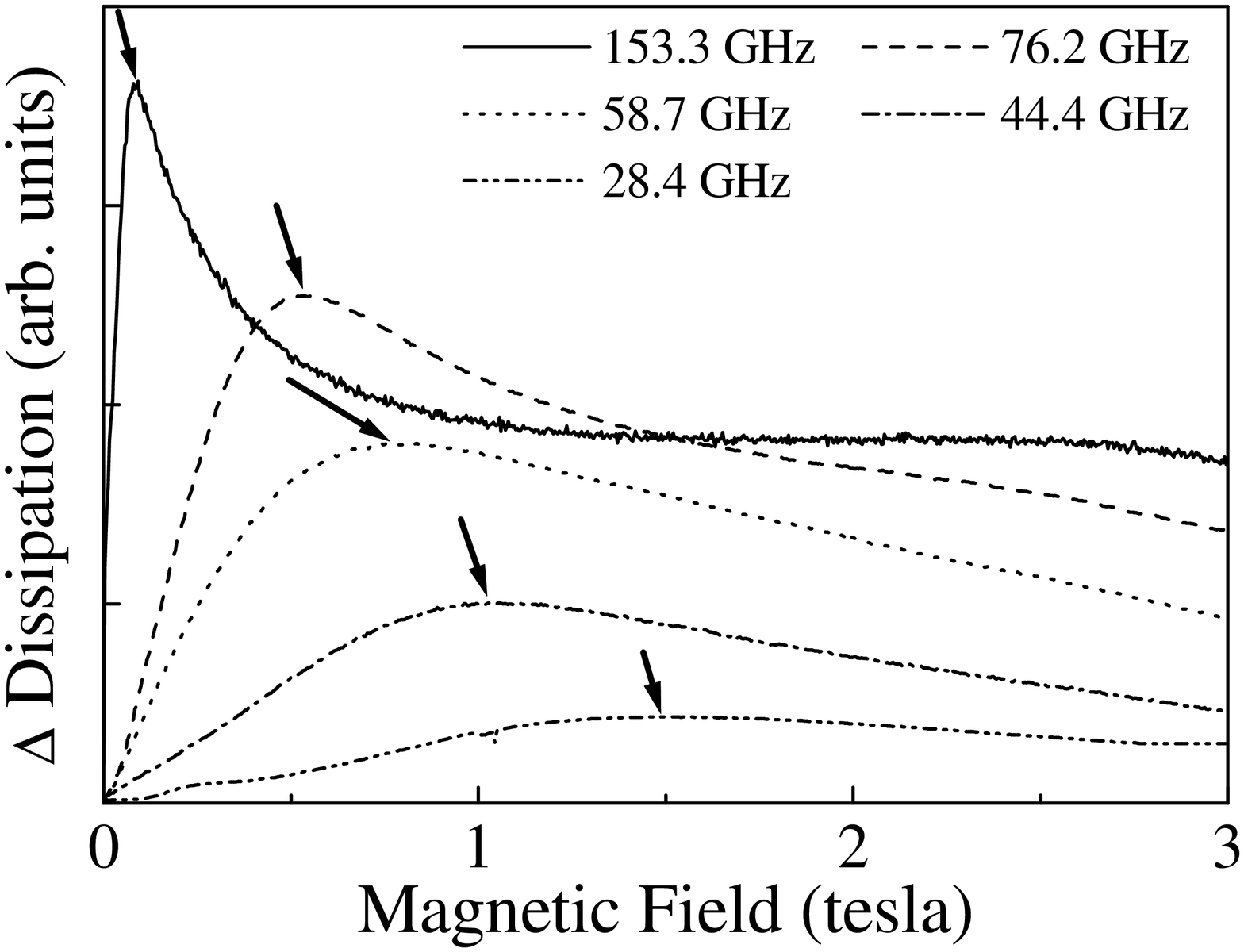,width=120mm}}
\bigskip\caption{M. Mola
{\em et al.}}

\end{figure}

\clearpage

\begin{figure}
\centerline{\epsfig{figure=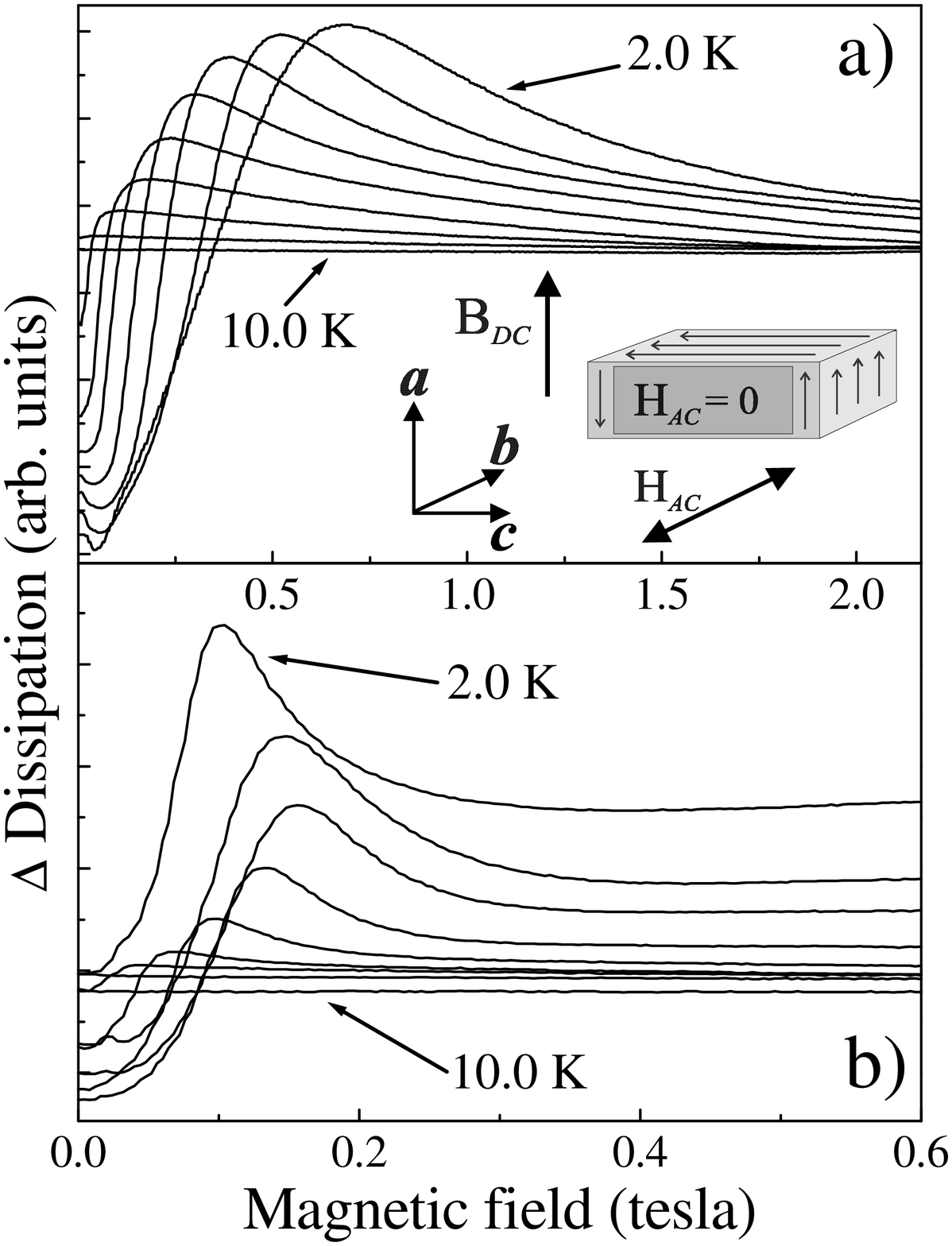,width=120mm}}
\bigskip\caption{M. Mola
{\em et al.}}

\end{figure}

\clearpage

\begin{figure}
\centerline{\epsfig{figure=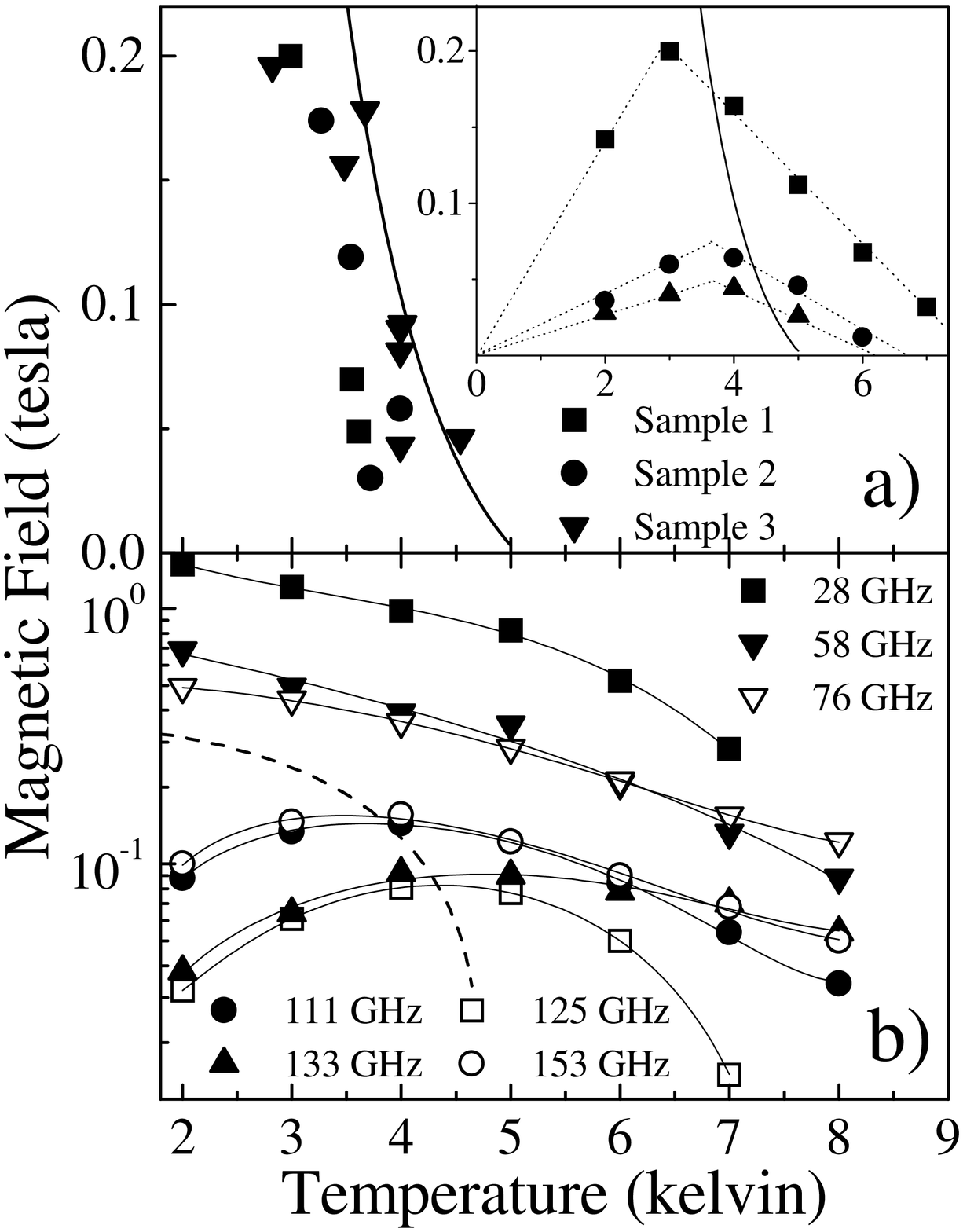,width=120mm}}
\bigskip\caption{M. Mola
{\em et al.}}

\end{figure}

\begin{figure}
\centerline{\epsfig{figure=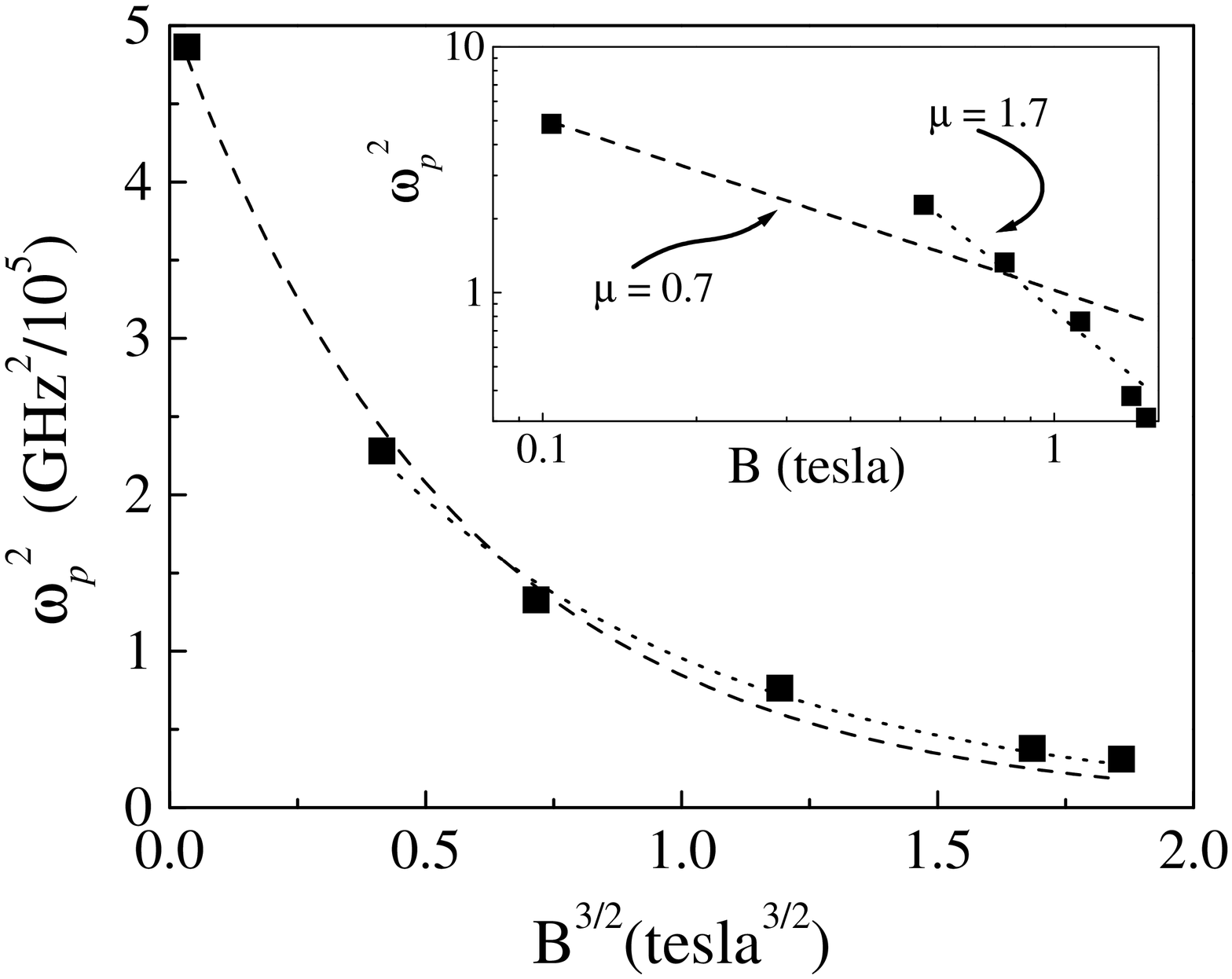,width=120mm}}
\bigskip\caption{M. Mola
{\em et al.}}

\end{figure}

\clearpage


\begin{references}
\bibitem{Bd}
We note that the decoupling field $B_D$ is quite different from
the dimensional crossover field, which represents the point at
which ordering of vortices changes over from a 3D line lattice to
a Q2D pancake vortex lattice. The former transition represents the
magnetic field range above which inter-layer transport ceases to
be dissipationless.
\end{references}
\end{document}